\documentclass[twocolumn,showpacs,preprintnumbers,amsmath,amssymb]{revtex4}

\usepackage{graphicx}
\usepackage{dcolumn}
\usepackage{bm}
\usepackage{amsfonts}

\newcommand{\ket}[1]{\mbox{$\mid \! #1 \, \rangle$}}

\newcommand{\bracket}[2]{\langle \, #1 \mid #2 \, \rangle}
\newcommand{\ketbra}[1]{\!\mid \! #1 \, \rangle\langle \, #1 \! \mid}
\newcommand{\crsp}{\,\widehat{=}\,}

\newcommand{\absl}[1]{\left| #1 \right|}
\newcommand{\dg}[1]{#1 \,^{\circ}}

\begin{document}


\title{Single Qubit Quantum Secret Sharing}

\author{Christian Schmid$^{1,2}$, Pavel Trojek$^{1,2}$, Harald Weinfurter$^{1,2}$, Mohamed Bourennane$^{3}$, Marek \.{Zukowski}$^{4}$ and Christian Kurtsiefer$^{5}$}

\affiliation{ $^{1}$Sektion Physik,
Ludwig-Maximilians-Universit{\"a}t, D-80797 M{\"u}nchen,
Germany\\
$^{2}$Max-Planck-Institut f{\"u}r Quantenoptik, D-85748 Garching, Germany\\
$^{3}$Department of Microelectronics and Information Technology,
Royal
Institute of Technology, SE-164 40 Kista, Sweden\\
$^{4}$Instytut Fizyki Teoretycznej i Astrofizyki, Uniwersytet
Gda\'{n}ski, PL-80-952 Gda\'{n}sk, Poland\\
$^{5}$Department of Physics, National University of Singapore,
Singapore 117 542, Singapore
}%

\date{December 29, 2004}

\begin{abstract}
We present a simple and practical protocol for the solution of a
secure multiparty communication task, the secret sharing,  and its
experimental realization. In this protocol, a secret message is
split among several parties in a way that its reconstruction
require the collaboration of the participating parties. In the
proposed scheme the parties solve the problem by a sequential
communication of a single qubit. Moreover we show that our scheme
is equivalent to the use of a multiparty entangled GHZ state but
easier to realize and better scalable in practical applications.
\end{abstract}

\maketitle


Splitting a secret message in way that a single person is not able
to reconstruct it is a common task in information processing and especially high security applications. 
Suppose e.g. that the launch sequence of a nuclear missile is
protected by a secret code, and it should be ensured that not a
single lunatic is able to activate it but at least two lunatics. A
solution for this  problem and its generalization including several
variations is provided by classical cryptography \cite{brucesch} and
is called secret sharing. It consists of a way of splitting the
message using mathematical algorithms and the distribution of the
resulting pieces to two or more legitimate users by classical
communication. However all ways of classical communication currently
used are susceptible to eavesdropping attacks. As the usage of
quantum resources can lead to unconditionally secure communication
(e.g. \cite{gisin, ekert}), a protocol introducing quantum
cryptography to secret sharing was proposed \cite{HARALD, qsecsh,
got, karl}. In this protocol a shared GHZ-state allows the
information splitting and the eavesdropper protection
simultaneously. But, due to lack of efficient multi-photon sources
an experimental demonstration of secret sharing is still missing.
Till now solely the principle feasibility of an experimental
realization using pseudo-GHZ states was shown \cite{tittel}.

Here we propose a protocol for $(N+1)$ parties in which only
sequential single qubit communication between them is used and
show its equivalence to the GHZ-protocol. As our protocol requires
only single qubits it is realizable with the
current state-of-the-art technologies and above all much more
scalable with respect to the number of participating parties.
These gains enabled the experimental demonstration of our protocol
for six parties. To our knowledge this is the first experimental
implementation of a full protocol for secret sharing and by far
the highest ever reported number of participants in any quantum
information processing task.


Let us first shortly describe the entanglement based protocol using
a GHZ state for secret sharing. Consider $(N+1)$ persons, each
having a particle from the maximally entangled $(N+1)$ particle
GHZ-state
\begin{equation}
\ket{GHZ}=\frac{1}{\sqrt{2}}\left(\ket{\underbrace{00\dots0}_{N+1}}+\ket{\underbrace{11\dots1}_{N+1}}\right).
\end{equation}
One of the parties, let's call him distributor, wants to
distribute a secret message among the remaining $N$ persons
(recipients) in a way that all of them 
have to cooperate in order to reconstruct the distributed message.
To achieve this task each participant performs a projection
measurement of his particle onto the eigenstates
$\ket{k_j,\phi_j}=1/\sqrt{2}(\ket{0}+k_j \exp(i\phi_j)\ket{1})$
($j=1,2,\dots,N+1$) of the operator
\begin{equation}
\widehat{\sigma}_j(\phi_j)=\sum_{k_j} k_j\ketbra{k_j,\phi_j},
\end{equation} where $k_j=\pm1$ denotes the local result in mode
$j$ for a preselected parameter $\phi_j$. The partners randomly
and independently choose between $\phi_j=0$ or $\pi/2$. The
correlation function for a $(N+1)$ particles GHZ state is defined
as the expectation value of the product of $(N+1)$ local results
and is therefore given by
\begin{equation}\label{eqn:ghzcorr}
E(\phi_j)=\langle\prod_j^{N+1}
\widehat{\sigma}_j(\phi_j)\rangle\;=\;\cos\left(\sum_j^{N+1}\phi_j\right).
 \end{equation}

After the measurement each recipient publicly announces her/his
choice of $\phi_j$, but keeps the result $k_j$ secret. By doing so
the distributor can decide when this procedure leads to perfect
(anti-)correlated results, i.e. when $|\cos(\sum_j^N\phi_j)|=1$,
which happens in half of the runs. In these instances each of the
recipients is able to infer the distributor's measurement result
$k_d$ if and only if he/she knows the measurement results $k_r$
($r=1,2,\dots,N$) of all the other recipients. Consequently the
cooperation of all the recipients is required and  any subset of
the parties has no information on the secret. For a security proof
of this scheme against eavesdropping attacks see
\cite{qsecsh,scarani}.

An equivalent $(N+1)$ party scheme (see fig. \ref{fig:setup}) for
the same task where only the sequential communication of a single
qubit is used, runs as follows.
\par
The distributor randomly prepares a qubit in one of the four
states $\ket{\pm x}, \ket{\pm y}$ of two mutually unbiased bases x
and y with
\begin{align}
\ket{\pm x}&=\frac{1}{\sqrt{2}}(\ket{0}\pm \ket{1})\\
\ket{\pm y}&=\frac{1}{\sqrt{2}}(\ket{0}\pm i \ket{1}).
\end{align}
Note that all these states are of the form
\begin{equation}
\ket{\chi}_i=\frac{1}{\sqrt{2}}\left(\ket{0}+e^{i\varphi_d}\ket{1}\right),
\end{equation}
where $\varphi_d$ is chosen to have one out of the four values
$\{0, \pi , \pi/2, 3\pi /2 \}$.
\begin{figure}
\includegraphics[scale=0.43,clip]{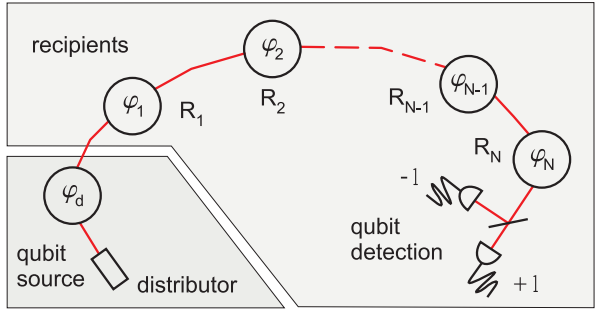}
\caption{\label{fig:setup}Scheme for $(N+1)$ party single qubit
secret sharing. The distributor prepares a qubit in an initial
state and acts on it with the phase operator
$\widehat{\sigma}(\varphi_d)$. Afterwards the qubit is
sequentially communicated from one recipient to another each
acting on it with $\widehat{\sigma}(\varphi_j)$ as well. The last
recipient performs finally a measurement of the qubit leading to
the result $\pm 1$. In half of the cases the phases add up such
that the preparation and the measurement are perfectly
(anti-)correlated.}
\end{figure}
During the protocol the qubit is then sequentially communicated
from recipient to recipient each acting on it with the unitary
phase operator
\begin{equation}
\widehat{\sigma}_j(\varphi_j)=\begin{cases} \ket{0}\rightarrow \ket{0}&\\
\ket{1}\rightarrow e^{i\varphi_j}\ket{1},
\end{cases}
\end{equation}
where $\varphi_j \in \{0, \pi , \pi/2, 3\pi /2 \}$ as well.
Therefore having passed all parties the qubit will end up in the
state
\begin{equation}
\ket{\chi}_f=\frac{1}{\sqrt{2}}\left(\ket{0}+e^{i(\varphi_d+\sum_j
\varphi_j)}\ket{1}\right).
\end{equation}
After this communication stage each participant divides his action
for every run into two classes: a class X corresponding to the
choice of $\varphi_j\in \{0, \pi\}$ and a class Y corresponding to
$\varphi_j\in \{\pi/2, 3\pi /2\}$. Following this classification
they inform the distributor about the class-affiliation of their
action for each run. Note that they keep the particular value of
$\varphi_j$ secret. This corresponds to the announcement of $\phi_j$
while keeping $k_r$ secret in the GHZ-scheme. The order in which the
recipients $R_j$ announce the class-affiliation is randomly
determined by the distributor. The last recipient $R_N$ finally
measures the received qubit in the x basis. Therefore for her/him it
suffices to choose only between $\varphi_N=0$ or $\varphi_N=\pi/2$
and keep the outcome $k_N$ of the measurement secret \cite{remark1}.
The probability that $R_N$ detects the state $\ket{+x}$ is given by
\begin{equation}
p_+(\varphi_d,\varphi_{1},\dots,\varphi_N)=\frac{1}{2}(1+\cos
(\varphi_d+\sum_j^N\varphi_{j})),
\end{equation}
whereas the probability to detect the state $\ket{-x}$ is
\begin{equation}
p_{-}(\varphi_d,\varphi_{1},\dots,\varphi_N)=\frac{1}{2}(1-\cos
(\varphi_d+\sum_j^N\varphi_{j})).
\end{equation}
So the expectation value of the measurement result is
\begin{multline}\label{eqn:singlecorr}
A(\varphi_d,\varphi_{1},\dots,\varphi_N)=p_+(\varphi_d,\varphi_{1},\dots,\varphi_N)\\
-p_-(\varphi_d,\varphi_{1},\dots,\varphi_N)=\cos(\varphi_d+\sum_j^N
\varphi_j).
\end{multline}

From the broadcasted class-affiliations of all introduced phase
shifts $\varphi_j$ the distributor is able to decide which runs
lead to perfect (anti-)correlations, means when
$|\cos(\varphi_d+\sum_j^N\varphi_j)|=1$, what happens in half of
the runs. We call this a valid run of the protocol. In these cases
each of the recipients is able to infer the distributor's choice
of $\varphi_d$ if and only if he/she knows the choice of
$\varphi_j$ of the other recipients. Consequently the
collaboration of all recipients is necessary.

By associating the particular value of $\varphi_d$ with "0" and
"1", say e.g. $\varphi_d \in \{0, \pi/2\} \crsp 0$ and $\varphi_d
\in \{\pi, 3\pi/2\} \crsp 1$, the parties are able to secretly
share a common bit string (key). This is possible as obviously the
required correlations based on local manipulation of relative
phases can equivalently be established by communicating a single
qubit instead of employing many entangled qubits of a GHZ-type
state; (compare equation \ref{eqn:ghzcorr} and
\ref{eqn:singlecorr}).

In order to ensure the security of the protocol against
eavesdropping or cheating \cite{remark2} the distributor arbitrarily
selects a certain number (might depend on the degree of security
requirements) of particular valid runs. For this subset the
correlations are publicly compared, again in a random order of the
recipients. The public comparison will reveal any eavesdropping or
cheating strategy. That can be easily seen from the following
intercept/resend eavesdropping attacks.

Imagine for instance the first recipient $R_1$ tries to infer the
secret without the help or the authorization of the remaining
participants by measuring the qubit sent by the distributor
\emph{before} acting on it with $\widehat{\sigma}_1(\varphi_1)$
and afterwards sending it ahead to the second recipient $R_2$. For
convenience, let us assume $R_1$ chooses for this measurement one
of the two protocol bases x or y. As the distributor applies
randomly one of four different phase shifts, the probability that
the state $\ket{\chi}_i$ is an eigenstate of the measurement
chosen by $R_1$ is 1/2. In the other half of the cases the
measurement result of $R_1$ will be completely random as it holds
that $\absl{\bracket{\pm y}{\pm x}}=\absl{\bracket{\pm x}{\pm
y}}=1/2$. This means that recipient $R_1$ gets no information
about the distributor's choice of $\varphi_d$. Furthermore this
cheating will cause an overall error of 25 \% in the correlations.
That's because if $R_1$ has chosen the wrong basis, the final
state of the qubit after all $(N+1)$ introduced phase shifts will
be of the form
\begin{equation}
\ket{\chi}_{f\prime}=\frac{1}{\sqrt{2}}\left(\ket{0}+e^{i\sum_{j=1}^N
\varphi_j} \ket{1}\right)
\end{equation}
instead of $\ket{\chi}_{f}$.

The state $\ket{\chi}_{f\prime}$ will, measured by the last
recipient $R_N$, give with probability 1/2 a result which is not
compatible to the expected correlations. The same situation an
eavesdropper is faced with, when applying such a strategy. The
usage of the bases x and y for an intercept/resend attack is
already the optimal one concerning the information gain on the
valid runs. One might only consider using the intermediate (or so
called Breidbart) basis $\ket{\pm
b}\frac{1}{\sqrt{2+\sqrt{2}}}(\ket{\pm x} + \ket{\pm
y})=\frac{1}{\sqrt{2}}(\ket{0}\pm e^{i \pi/4}\ket{1})$ which gives
the eavesdropper maximum information on all exchanged bits
\cite{hutek}. But even here the error rate goes necessarily up to
25 \%. The security of the presented protocol against a general
eavesdropping attack follows from the proven security (see for
detail \cite{gisin}) of the well known BB84 protocol \cite{bb84}.
Each communication step between two successive parties can be
regarded as a BB84 protocol using the bases x and y. Any set of
dishonest parties in our scheme can be viewed as an eavesdropper
in BB84 protocol.

The presented protocol was experimentally implemented for six
(5+1) parties, thus clearly showing the practicality and
user-friendliness of the scheme.

We encoded the qubit of the protocol in a single photon where the
basis states $\ket{0}$ and $\ket{1}$ are represented by the
polarization states of the photon $\ket{H}$ and $\ket{V}$
respectively, corresponding to horizontal (H) and vertical (V)
linear polarization. The single photons were provided by a
heralded single photon source. The setup is shown in
Fig.~\ref{fig:setup}.
\begin{figure}
\includegraphics[scale=0.37,clip]{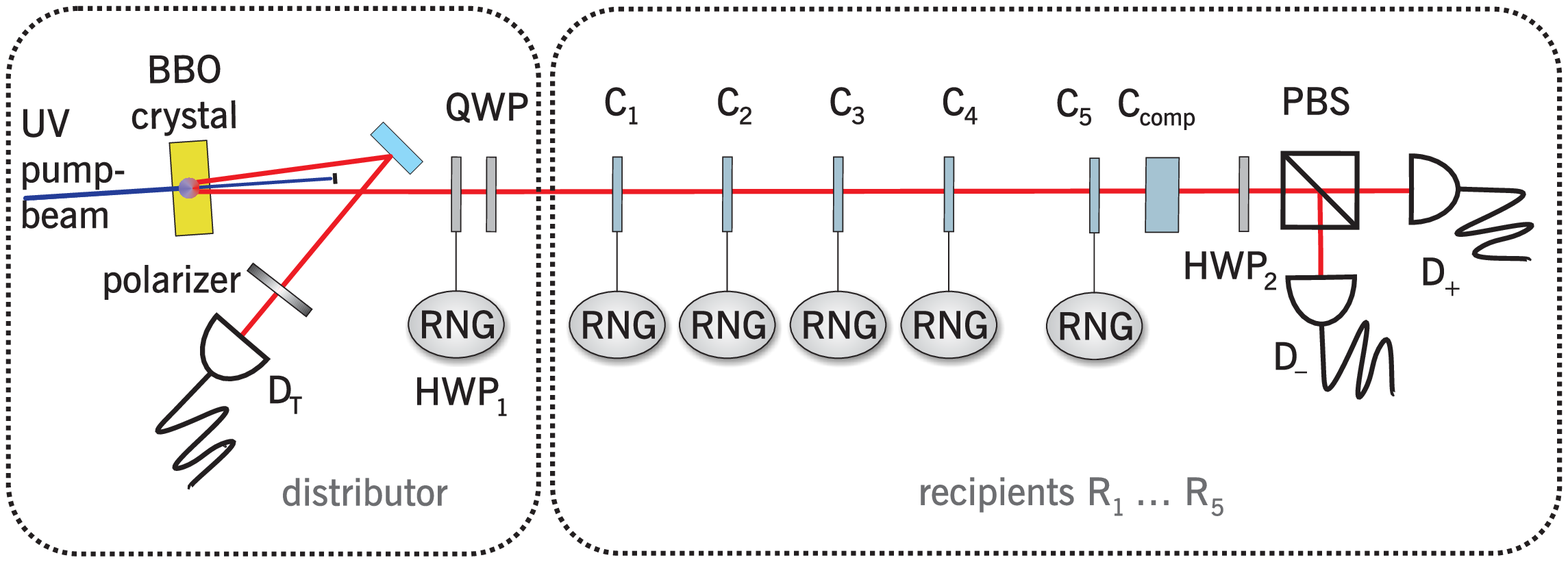}
\caption{\label{fig:setup}Setup for single qubit secret sharing.
Pairs of orthogonally polarized photons are generated via a type
II SPDC process in a BBO crystal. The detection of one photon from
the pair by D$_{\mathrm{T}}$ heralds the existence of the other
one used for the performance of the protocol. The initial
polarization state is prepared by the distributor by a polarizer
in front of the trigger detector and a half- and quarter wave
plate (HWP$_1$, QWP). Each of the recipients ($R_1 \dots R_5$)
introduces one out of four phase shifts according to a number from
a pseudo random number generator (RNG) by the rotation of YVO$_4$
crystals (C$_1 \dots$C$_5$). The last party analyzes additionally
the resulting polarization state of the photon with a half-wave
plate (HWP$_2$) and a polarizing beam splitter. }
\end{figure}
A pair of photons is created via a spontaneous parametric down
conversion (SPDC) process. As the photons of a pair are strongly
correlated in time the detection of one photon in D$_{\mathrm{T}}$
heralds the existence of the other one which is used for the
protocol. Thus from a coincidence detection between
D$_{\mathrm{T}}$ and D$_+$/D$_-$ within a chosen time window of 4
ns we assume the communication of a single photon only. For this
coincidence time window and singlecount rates of about 70000
$\mathrm{s}^{-1}$ in D$_+$/D$_-$ accidental coincidences were
negligible. The SPDC process was run by pumping a 2 mm long
$\beta$-barium borate (BBO) crystal with a blue single mode laser
diode (402.5 nm) at an optical output power of 10 mW. Type-II
phase matching was used at the degenerate case leading to pairs of
orthogonally polarized photons at a wavelength of $\lambda=805$ nm
($\Delta \lambda \approx 6$ nm).

In order to prepare the initial polarization state a polarizer
transmitting vertically polarized photons was put in front of the
trigger detector D$_{\mathrm{T}}$ ensuring that only horizontally
polarized photons can lead to a coincidence detection. The
distributor was equipped with a motorized half-wave plate
(HWP$_1$) followed by quarter-wave plate (QWP) at an angle of
$\dg{45}$. By rotation of HWP$_1$ to the angles $\dg{0},\dg{45}$
and $\dg{22.5},\dg{-22.5}$ he could transform the horizontally
polarized photons coming from the source to $\ket{\pm y}$ and
$\ket{\pm x}$. This corresponds to applying the phase-shifts
$\varphi_d\in\{\pi/2,3\pi/2\}$ and $\varphi_d\in\{0,\pi\}$
respectively. As the phase-shifts of the recipients had to be
applied independently from the incoming polarization state the
usage of standard wave plates was not possible. Therefore the
unitary phase operator was implemented using birefringent uniaxial
200 $\mu$m thick Yttrium Vanadate (YVO$_4$) crystals (C$_i$). The
crystals were cut such that their optic axis lies parallel to the
surface and aligned that H and V polarization states correspond to
their normal modes. Therefore by rotating the crystals along the
optic axis for a certain angle a specific relative phase shift was
applied independent from the incoming polarization state. An
additional YVO$_4$ crystal (C$_{comp}$, 1000 $\mu$m thick) was
used to compensate for dispersion effects. The last party
performed the projection measurement using a half-wave plate
(HWP$_2$) at an angle of $\dg{22.5}$ followed by polarizing
beam-splitter (PBS). The photons were detected at D$_+$/D$_-$ and
D$_{\mathrm{T}}$ by passively quenched silicon avalanche photo
diodes (Si-APD) with an efficiency of about 35 \%.

\begin{table}[t]
\begin{ruledtabular}
\begin{tabular}{l|c c c c c}
   & $z_{total}$ & $z_{one}$ & $z_{raw}$ & $z_{val}$ & QBER [\%] \\
  \hline
  $\ket{\pm x}$ & 27501 & 9814 & 883 & 452 & $25.22 \pm 2.04$\\
  $\ket{\pm y}$ & 24993 & 9188 & 784 & 409 & $30.32 \pm 2.27$\\
  $\ket{\pm b}$ & 38174 & 13706 & 1137 & 588 & $30.27 \pm 1.89$\\
\end{tabular}
\end{ruledtabular}
\caption{\label{tab:res}Results of the simulation of an
intercept/resend eavesdropping strategy in the protocol- and
intermediate basis. The attack was done by inserting a polarizer
between the distributor and the first recipient. In each case the
quantum bit error rate (QBER) rises up to more than 25 \% and by
this blows the eavesdropper's cover. }
\end{table}

The protocol was repeated $z_{total}=25000$ times. One run
consisted of rotating the crystals and opening the detectors for a
collection time window $\tau=200\,\mu$s what took together about 1
s. Each crystal was thereby driven by a motor to one of four
different positions given by a pseudo random number. This means
the application of one of the four phase shifts at random by each
party. Out of $z_{total}$ only $z_{one}=9125$ times exactly one
photon was detected at D$_{\mathrm{T}}$ within $\tau$ due to
poissonian photon-counting statistics. In these runs a coincidence
detection happened $z_{raw}=2107$ times which provided us with the
raw key. From this we extracted $z_{val}=982$ valid runs where
$|\cos(\sum_j^N\varphi_j)|=1$ (506 times
$\cos(\sum_j^N\varphi_j)=1$ and 476 times
$\cos(\sum_j^N\varphi_j)=-1$ ) with a quantum bit error rate
(QBER) of $2.34 \pm 0.48$ \%.

In order to show that the QBER increases significantly by an
eavesdropping attack we simulated an intercept/resend strategy by
inserting a polarizer between the distributor and the first
recipient. The attack was done in the protocol bases $\ket{\pm x},
\ket{\pm y}$ as well as in the intermediate basis $\ket{\pm b}$.
For the latter two the polarizer was additionally sandwiched by
two quarter-wave plates. The angular settings (1st QWP, polarizer,
2nd QWP) were $\{\dg{45},\dg{0},\dg{-45}\}$ and
$\{\dg{-45},\dg{22.5},\dg{45}\}$. For every choice of the basis
the QBER went up to at least 25 \% (or even higher due to other
experimental imperfections). The results are summarized in
Table~\ref{tab:res}.

In summary, we introduced a new scheme for solving the multi-party
communication task of secret sharing. Unlike other schemes
employing multi-particle entangled states our protocol uses only
the sequential communication of a single qubit. As single qubit
operations using linear optical elements and the analysis of
photon polarization states are quite well accomplishable with
present day technology, we were therefore able to present a first
experimental demonstration of the protocol for six parties. This
is to our knowledge the highest number of actively performing
parties in a quantum protocol ever implemented so far, and the
first ever experimental implementation of a full quantum secret
sharing protocol. We also simulated an eavesdropping
intercept/resend attack and by this showed the resistance of the
protocol against such kind of strategies because of a
significantly increasing error rate. In principle we see no
experimental barrier to extend the performed protocol to even
significantly higher number of participants. The achieved key
exchange rate could be easily increased  by using fast
electro-optical phase modulators. Also the use of weak coherent
pulses of light containing much less than one photon on average,
instead of a heralded single photon source, is  possible and might
further reduce the experimental effort. However, this would be at
the expense of the concept of communicating strictly one qubit and
can be also disadvantageous for the practical performance of the
protocol \cite{sanders, lut}. While we have realized our secret
sharing protocol using photons and polarization encoding,
alternative schemes, like proposed or realized in BB84-type
protocols can be adopted as well. One might think of other forms
of information encoding, higher multilevel, or continuous
variables. Finally we stress that by showing that our approach is
equivalent to the use of a many qubit GHZ state we opened the door
to the possible application of this method in other generic multi
party communication tasks.

M.\.{Z}. is supported by an FNP Profesorial Subsidy, and MNiI Grant
1 P03B 04927. The work is a part of MNiI/DAAD collaboration program
and was furthermore supported by German DFG and BMBF, the Bavarian
high-tech initiative, Swedish Research Council (VR), and the
European Commission through the IST FET QIPC RamboQ.


\begin{thebibliography}{15}
\bibitem{remark1}Alternatively it is of course also possible
to choose $\varphi_N$ out of four values and announce in addition to
the class-affiliation the value of $k_N$. The parties should only
agree in the beginning which variation of the protocol they prefer.
\bibitem{remark2}By eavesdropping we
refer to an attack from a person which is not officially
participating
   in the protocol whereas by cheating we refer to an attack from a participant.
\end{thebibliography}


\end{document}